\documentclass[a4paper,11pt]{article}
\usepackage{jheppub} 
\usepackage{lineno}

\usepackage[utf8]{inputenc}
\usepackage{amsmath}
\usepackage{amsfonts}
\usepackage{graphicx}
 
\newcommand{\dd}{\mathrm{d}}
\newcommand{\Lag}{\mathcal{L}}
\newcommand{\EoM}{\mathcal{E}}

\newcommand{\iN}{\mathrm{i}}

\newcommand{\mpl}{M_\mathrm{Pl}}
\newcommand\cutoff{\Lambda_{{\scriptscriptstyle(3)}}}
\newcommand{\aECG}{\alpha}
\newcommand{\bECG}{\beta}

\newcommand\sphint[1]{\int_{\mathbb{S}^2}\,\, #1 \,\, \dd\Omega}
\newcommand\lm{{\ell m}}
\newcommand\lmp{{\ell'm'}}
\newcommand\Ylm{Y_\lm}
\newcommand\Xlm{X_\lm}

\newcommand\p{\prime}
\newcommand\pp{\p\p}
\newcommand\ppp{\p\p\p}
\newcommand\pppp{\p\p\p\p}

\title{On the physical viability of black hole solutions in Einsteinian Cubic Gravity and its generalisations
}

\author[1]{Jose Beltr\'an Jim\'enez}
\affiliation[1]{Departamento de F\'isica Fundamental and IUFFyM, Universidad de Salamanca, E-37008 Salamanca, Spain}

\author[2]{and Alejandro Jim\'enez-Cano}
\affiliation[2]{Laboratory of Theoretical Physics, Institute of Physics, University of Tartu, W. Ostwaldi 1, 50411 Tartu, Estonia\\$\,$}

\emailAdd{jose.beltran@usal.es}
\emailAdd{alejandro.jimenez.cano@ut.ee}

\abstract{
In this note, we discuss the pathological nature of black holes in Einsteinian Cubic gravity and its extensions. We compute the equations for the odd perturbations and show how spherically symmetric solutions that asymptotically approach a maximally symmetric space (Minkowski, de Sitter or anti-de Sitter) are associated to having an asymptotically degenerate principal part of the equations. We use these results to argue that the encountered problems will be generic for any other cubic or higher-order with a reduced linear spectrum around maximally symmetric spaces except the well-known healthy case of $f(R)$. We highlight that these pathologies are only alarming when the theory is regarded as a complete theory, but not when considered as a perturbative correction to GR (as in e.g. the effective field theory framework) since the the low-energy physics remains safe from them. Our results thus cast doubts on possible resolutions of the singularities or non-perturbative effects on horizons based on these theories.
}

\begin{document}

\maketitle
\flushbottom

\section{Introduction}

General Relativity provides an exquisite description of most gravitational phenomena ranging from the cosmological evolution to the detected signals of black hole mergers. Despite its uncanny success, it is generally accepted that it can only be the low-energy effective field theory of some more fundamental theory of gravity. With this philosophy in mind, one expects to have corrections with higher powers of the curvature to the Einstein-Hilbert action. These higher-order terms generically give rise to higher-order equations of motion with the associated ghost modes vindicated by Ostrogradski's construction. Regarding these higher-order terms as perturbative, the mass of these ghosts would be above the cut-off of the theory so they are harmless for the low-energy phenomenology. In arbitrary dimension, it is possible to have higher-order curvature terms that do not introduce new modes so they could be considered in the non-perturbative regime. According to Lovelock's theorem \cite{Lovelock1970,Lovelock1971}, gravity theories lying outside the Lovelock class present higher than second-order time derivatives of the metric in any $3+1$ splitting, hence Ostrogradski theorem applies and unstable ghostly degrees of freedom (dof's) are expected around generic backgrounds \cite{woodard2019theorem, Woodard:2006nt}. In four dimensions, Lovelock's theorem only allows to have the cosmological constant and the Ricci scalar, while the Gauss-Bonnet term $G$ is topological. Known ghost-free exceptions with additional dof's are those based on non-linear extensions of the Lovelock terms such as $f(R)$ or $f(G)$, that also exist in four dimensions. Theories beyond these classes are therefore prone to having ghost-like instabilities so physical solutions where higher-order terms are relevant necessarily exhibit a pathological behaviour. As commented above, such terms can only make sense in a perturbative regime where the effects are also perturbative. 

Despite the restrictive character of the Lovelock's theorem, there have been some attempts at obtaining black hole solutions different from GR that rely on higher-order curvature terms \cite{Hennigar2016, BuenoBH, Bueno2017BHstabil, Lessa2023}. These scenarios aim at getting around the mentioned problems by carefully tuning coefficients in the action so as to have a reduced linear spectrum around specific backgrounds. In this paper we will focus on a particular family of cubic corrections of GR containing two subcases with some remarkable properties. One is the so-called \emph{Einsteinian Cubic Gravity} (ECG) \cite{Oliva:2010eb,Myers:2010ru,BuenoCanoECG}. For our purposes, the most interesting property of this theory is that its spectrum around maximally symmetric spacetimes of arbitrary dimension corresponds to that of GR. In four dimensions, it is possible to construct an improved version of ECG which possess the GR spectrum around any Friedman-Lema\^itre-Robertson-Walker spacetime, not only maximally symmetric ones. The latter is known in the literature as \emph{Cosmological Einsteinian Cubic Gravity} (CECG), and is also a particular case of the theories that we explore in the present work.

These theories do not belong to the Lovelock class and, hence, they present ghostly dof's. However, as mentioned previously, one can find that the kinetic terms of the extra dof's vanish for some specific background solutions, so the only propagating dof's are those of GR (the graviton). This typically signals the so-called \emph{strong coupling problem}, namely, that the disappearing dof's are actually strongly coupled for arbitrarily small deviations of the considered background. The exact solution cannot be obtained by perturbatively solving the equations of motion, and the linear spectrum analysis does not capture all the properties of the evolution of such deviations. This pathological behaviour usually leads to unstable backgrounds as shown for the cosmological case in e.g. \cite{BeltranJC2021}. The present work is a continuation of \cite{BeltranJC2021} aimed at extending the analysis to spherically symmetric backgrounds. We will consider a class of theories that interpolate between the CECG and ECG and, additionally, we will include another free parameter in the action controlling a cubic deviation from the previous ones. 

For the reasons mentioned above, we will focus on spherically symmetric solutions in these theories. Moreover, in the perturbation analysis, we will concentrate on odd-parity modes with $\ell>1$, for simplicity. First, we will show that the equations of motion for these two modes are third-order in time derivatives, signalling the presence of 3 instead of 2 dof's, one of them coming from the extra order in time derivatives, which is expected to render an unstable evolution. We will also prove that for asymptotically flat spaces, the order in time derivatives of the system abruptly changes (as in \cite{BeltranJC2021}), indicating that the missing dof's exhibit a strongly coupled behaviour.

Before closing this introduction, we would like to mention that during the realization of this work, the article \cite{DeFelice2023} was released covering similar topics. As we will discuss in more detail in the following, our results are complementary and we provide an alternative understanding of their results. Our discussion makes it clear that the pathologies found in these theories are actually not surprising and they could be anticipated because they are a general feature of higher-order curvature theories outside the Lovelock class. We also discuss how considering these theories as effective field theories (EFTs) allow to bypass these pathologies, an approach that has also been put forward in the recent work \cite{Bueno2023}, where the authors insist on the viability of the solutions only if higher-order terms in the effective field theory (EFT) remain perturbative and the scales to be explored are well below the cut-off. In this work we will try to complement the latter by connecting these issues with the strong coupling problem mentioned above.

The paper is organised as follows: In Section \ref{sec:higherordertheories}, we discuss the generic problems that arise in higher-order gravity theories due to the presence of extra ghostly dof's. In Section \ref{sec:ECG} we introduce a family of cubic theories that extends ECG and CECG. The main section of this text is Section \ref{sec:stab}, in which we focus on the odd-parity equations, and discuss the present pathologies associated to the extra dof's. Finally, in Section \ref{sec:conclusions}, we collect the most relevant ideas of this work and argue that these problems will generically appear in other higher-order gravity theories. Additionally, we provide some appendices giving more details on our derivations at the end of the manuscript.

~

\noindent \textbf{Conventions}:  We will use the same conventions as in the previous work \cite{BeltranJC2021}, namely, for the metric we take the mostly plus signature and units are chosen so that  $c=1$ and $8\pi G_{\text{N}} = \mpl^{-2}$. For symmetrization and antisymmetrization of indices we use, respectively $(\mu_1...\mu_n)$ and $[\mu_1...\mu_n]$, which include a symmetry factor $\frac{1}{n!}$. Moreover, we define the curvature tensors following \cite{Wald1984}, namely for the Riemann tensor we have $R_{\mu\nu\rho}{}^\lambda \equiv - 2 \big( \partial_{[\mu}\Gamma^\lambda_{\nu]\rho} + \Gamma^\lambda_{[\mu|\sigma} \Gamma^\sigma_{|\nu]\rho} \big)$ and the Ricci tensor is given by $R_{\mu\nu}\equiv R_{\mu\lambda\nu}{}^\lambda$.

\section{Reduced spectrum and strong coupling}
\label{sec:higherordertheories}

The original ECG has the property of sharing the same linear spectrum as GR on maximally symmetric backgrounds and this property was then showed to be shared by a more general class of four-dimensional theories dubbed Generalised Quasi-topological gravities (GQT). All these theories are based on actions that contain cubic and higher-order terms in curvature and, thus, they generically lie outside the Lovelock class of theories that maintain the second-order nature of the equations of motion. In particular, in four dimensions, the Lovelock's theorem guarantees that all these theories will have higher-order equations of motion and, therefore, they will contain a ghostly mode.\footnote{An exception is of course the class of $f(R)$ theories that are equivalent to GR plus a healthy scalar field.} More specifically, the generic theory will contain eight dof's corresponding to a massless spin-2, a ghostly massive spin-2 (assuming that the usual graviton is required to be healthy) and a scalar.
This reasoning shows that the GQT theories will have more dof's than GR so that the property of exhibiting the same number of propagating dof's around maximally symmetric backgrounds actually suggests that these are strongly coupled and, consequently, their stability is compromised. We can be more explicit on this. Let us consider an arbitrary background $\bar{g}_{\mu\nu}$ and consider perturbations around it that we will collectively denote $\Psi_i$. The indices $i,j...$ stand for whatever quantum numbers that are associated to the physical modes, i.e., after solving for the constraints. This procedure might give rise to spatial non-localities, but this will not be important for our argument. For the physical modes, then the linear equations will adopt the schematic form
\begin{equation}
\mathcal{G}^{\mu\nu\rho\sigma ij}\partial_\mu\partial_\nu\partial_\rho\partial_\sigma\Psi_j+\mathcal{F}^i=0\,,
\end{equation}
where $\mathcal{F}^i$ includes all the terms with derivatives of order lower than four. The super-metric $\mathcal{G}^{\mu\nu\rho\sigma ij}$ will govern the additional dof's with respect to GR and depends on the background solution under consideration. In particular, backgrounds with a reduced spectrum (if any) will correspond to having a degenerate $\mathcal{G}^{\mu\nu\rho\sigma ij}$. For some theories, this can happen for an arbitrary background, thus reflecting that those particular theories do not propagate all 8 dof's. This will happen for theories such as $f(R)$ or $f(G)$, being $G$ the Gauss-Bonnet invariant. A problem arises when the reduced linearised spectrum only occurs around specific backgrounds. This is the case for instance for $f(G)$ theories where Minkowski exhibits the same spectrum as GR. When this happens, the corresponding background does not provide a physically viable solution because it corresponds to a singular surface (either in phase or configuration space) where the principal part of the equations, characterised by $\mathcal{G}^{\mu\nu\rho\sigma ij}$, is degenerate. The problem with such solutions can be seen from two perspectives. On the one hand, this means that perturbations around such a background will be strongly coupled because the couplings of the canonically normalised fields will diverge. On the other hand, if we start from an arbitrary configuration in phase space, the evolution will never arrive at the solution precisely because it corresponds to a singular surface that can be interpreted as a local separatrix (see e.g. \cite{BeltranJC2021}). The discussed strong coupling problem is generic, but for the sake of concreteness, we will restrict our analysis to a triparametric family of cubic theories of gravity (which includes ECG and CECG) and show explicitly how the strong coupling arises around some spherically symmetric solutions.

However, before proceeding further it is important to mention that these problems are trivially avoided by interpreting these theories in the realm of EFTs (see e.g. \cite{deRham:2020ejn}). We will not perform a detailed analysis of this, but simply discuss the main points. In that framework, the higher-order curvature terms are to be regarded as perturbative corrections to the Einstein-Hilbert term and the additional dof's associated to the higher-order derivative terms are never excited within the regime of validity of the theory. This is not only natural, but it is tightly linked to the very EFT philosophy. A precise discussion on how this applies to the specific case of ECG has been put forward in \cite{Bueno2023}. In particular, the instabilities found in \cite{DeFelice2023} were shown to lie beyond the regime of validity of the EFT so they are harmless. In this work, we instead want to stress that, taking the cubic theories as full theories on their own and not as EFTs, the spherically symmetric solutions corresponding to non-perturbative modifications of the GR solutions (e.g. to regularise the singularity, affect the horizon in a non-perturbative manner, etc.) are not physically viable because of the mentioned problems. More specifically, we want to emphasise that tuning the action parameters to eliminate dof's from the linear spectrum around specific backgrounds is a physically ill-defined way of defining the theory.

\section{Einsteinian Cubic Gravity and cubic extensions} \label{sec:ECG}

Let us now proceed with some explicit computations for the proxy theory that we will consider. Following the notation in \cite{ArciniegaCECG}, we introduce the cubic combinations
\begin{align}
    \mathcal{P}&:= 
    12 R_\mu{}^\rho{}_\nu{}^\sigma R_\rho{}^\tau{}_\sigma{}^\eta R_\tau{}^\mu{}_\eta{}^\nu  
    +R_{\mu\nu}{}^{\rho\sigma} R_{\rho\sigma}{}^{\tau\eta} R_{\tau\eta}{}^{\mu\nu}
    -12R^{\mu\nu}R^{\rho\sigma}R_{\mu\rho\nu\sigma}+8 R_\mu{}^\nu R_\nu{}^\rho R_\rho{}^\mu\,,\\
    \mathcal{C}&:= R^{\mu\nu}R_\mu{}^{\rho\sigma \tau}R_{\nu\rho\sigma\tau} - \frac{1}{4} R R_{\mu\nu\rho\sigma}R^{\mu\nu\rho\sigma} - 2 R^{\mu\nu}R^{\rho\sigma}R_{\mu\rho\nu\sigma} + \frac{1}{2} R R_{\mu\nu}R^{\mu\nu}\,,
\end{align}
in terms of which we will write the theory that we will consider and that is described by the following three-parametric cubic correction to the Einstein-Hilbert action:
\begin{equation} \label{eq:action}
    S = \int \dd^4x \sqrt{|g|} \left[- \Lambda_0 + \frac{\mpl^2}{2} \left( R + \frac{\bECG}{\cutoff^4} \Lag_{\text{cub}} \right)\right]\,,
\end{equation}
where the first term corresponds to the cosmological constant, $\bECG$ is a dimensionless real parameter, $\cutoff$ is a mass scale and $\Lag_{\text{cub}}$ is a cubic polynomial in the Riemann tensor given by
\begin{align}
    \Lag_{\text{cub}} & := - \frac{1}{8} (\mathcal{P} - 8\aECG \mathcal{C})-\frac{1}{8}(\gamma-1)R_{\mu\nu}{}^{\rho\sigma} R_{\rho\sigma}{}^{\tau\eta} R_{\tau\eta}{}^{\mu\nu}\\
    &= - \frac{1}{8} \Big(
    12 R_\mu{}^\rho{}_\nu{}^\sigma R_\rho{}^\tau{}_\sigma{}^\eta R_\tau{}^\mu{}_\eta{}^\nu  
    +\gamma R_{\mu\nu}{}^{\rho\sigma} R_{\rho\sigma}{}^{\tau\eta} R_{\tau\eta}{}^{\mu\nu}
    +8 R_\mu{}^\nu R_\nu{}^\rho R_\rho{}^\mu\nonumber \\
    & \qquad \qquad
    +2 \aECG R R_{\mu\nu\rho\sigma}R^{\mu\nu\rho\sigma}
    -8\aECG R^{\mu\nu}R_\mu{}^{\rho\sigma \tau}R_{\nu\rho\sigma\tau}\nonumber\\
    &\qquad\qquad + 4(4\aECG-3) R^{\mu\nu}R^{\rho\sigma}R_{\mu\rho\nu\sigma}
    -4\aECG R R_{\mu\nu}R^{\mu\nu}
    \Big)\,.
\end{align}
Here, $\aECG$ and $\gamma$ are dimensionless real parameters. For $\gamma=1$, the parameter $\aECG$ interpolates between the original formulation of ECG \cite{BuenoCanoECG} ($\aECG=0$) and CECG \cite{ArciniegaCECG} ($\aECG =1$).\footnote{
        For $\gamma=\aECG=1$, the cubic Lagrangian $\Lag_{\text{cub}}$ coincides with the combination called $\mathcal{R}_{(3)}$ in \cite{ArciniegaGeoInflation}.}
The parameter $\gamma$ was included to study deviations of $\mathcal{P}$ other than the one given by $\mathcal{C}$. For the rest of this paper we assume $\bECG\neq 0$.\footnote{
    From the EFT perspective, the term with $\gamma$ can be removed (i.e., we can take $\gamma=1$) since, ECG is the only possible dimension-six invariant up to field redefinitions in vacuum. However, in our analysis, this term plays a non-trivial role because we are considering \eqref{eq:action} as a full theory, not as an EFT.
}

If we perturb the metric as $g_{\mu\nu}=\bar{g}_{\mu\nu}+h_{\mu\nu}$ around a maximally symmetric background $\bar{g}_{\mu\nu}$ with $\bar{R}_{\mu\nu\rho\lambda}=\mathcal{R}(\bar{g}_{\mu\rho}\bar{g}_{\nu\lambda}-\bar{g}_{\mu\lambda}\bar{g}_{\nu\rho})$, where $\mathcal{R}$ is a real parameter with mass dimension 2, the background and linear equations corresponding to \eqref{eq:action}  are respectively,
\begin{align}
    0&= 2\mpl^{-2}\Lambda_0 - 6 \mathcal{R} -3 (\gamma-3)\frac{\bECG}{\cutoff^4} \mathcal{R}^3 \,,\label{eq:bgeqMaxSym}\\
    0&=2 \left[\mpl^{-2}\Lambda_0 - 2 \mathcal{R}  + 6(\gamma-1) \frac{\bECG}{\cutoff^4} \mathcal{R}^3\right] h_{\mu\nu} 
    + \left[\mathcal{R} - \frac{3}{2} (\gamma+1) \frac{\bECG}{\cutoff^4} \mathcal{R}^3\right] g_{\mu\nu} h \nonumber\\
    &\quad- \left[1 - \frac{3}{2} (5 \gamma-3) \frac{\bECG}{\cutoff^4} \mathcal{R}^2\right] \bar{\nabla}_{\mu}\bar{\nabla}_{\nu}h 
    - \left[ 1 + \frac{3}{2} (9 \gamma-11) \frac{\bECG}{\cutoff^4} \mathcal{R}^2\right] \bar{\square} h_{\mu\nu} \nonumber\\
    &\quad+ \left[2 + 3 (5 \gamma -7) \frac{\bECG}{\cutoff^4} \mathcal{R}^2\right] \bar{\nabla}_{(\mu|}\bar{\nabla}_{\rho}h_{|\nu)}{}^{\rho} 
   - \left[1  + \frac{3}{2} (3 \gamma -5) \frac{\bECG}{\cutoff^4} \mathcal{R}^2\right] g_{\mu\nu} \bar{\nabla}_{\rho}\bar{\nabla}_{\lambda}h^{\rho\lambda} \nonumber\\
    &\quad+ \left[1 +\frac{3}{2} (3\gamma -5) \frac{\bECG}{\cutoff^4} \mathcal{R}^2\right] g_{\mu\nu} \bar{\square} h \nonumber\\
    &\quad+ 3 (\gamma-1)\frac{\bECG}{\cutoff^4} \mathcal{R} \Big( \bar{\nabla}_{\mu}\bar{\nabla}_{\nu}\bar{\nabla}_{\rho}\bar{\nabla}_{\lambda}h^{\rho\lambda} -  2 \bar{\square}\bar{\nabla}_{(\mu|} \bar{\nabla}_{\rho}h_{|\nu)}{}^{\rho}  +  \bar{\square}^2 h_{\mu\nu}
    \Big)  \,, \label{eq:pertMaxSym}
\end{align}
where $h:= \bar{g}^{\mu\nu}h_{\mu\nu}$, $\bar{\square}:=\bar{g}^{\mu\nu}\bar{\nabla}_\mu\bar{\nabla}_\nu$ and indices are raised with the background inverse metric. The background equation, for the particular case $\gamma=1$, corresponds to the embedding equation presented in \cite{BuenoCanoECG} restricted to 4 dimensions. Observe that for $\gamma =3$, we recover the GR solution $\mathcal{R}=\Lambda_0/3\mpl^2$. The cubic terms in \eqref{eq:action} have two effects in the perturbed equation: they introduce corrections to the second-order terms with powers of the curvature parameter $\mathcal{R}$ and, in addition to this, new terms with 4th-order derivatives of the perturbation are generated. All of these higher-order terms have a common factor in front proportional to both the curvature parameter and the combination $\gamma -1$.  Observe that $\aECG$ is absent in this equation. In \eqref{eq:pertMaxSym}, we already sense that the Ostrogradski ghosts due to the higher-order time derivatives are fully active if $\gamma \neq 1$. For $\gamma =1$ there is an abrupt change in the derivative content and, consequently, in the behaviour of the system of differential equations. However, this is only an accident that occurs due to the highly symmetric background under consideration, but it is expected to disappear as we reduce the symmetries of the background. This is precisely what we intend to investigate in the subsequent sections with the corresponding implications of this discontinuity in the number of propagating dof's. We can however already anticipate that spherically symmetric solutions that approach a maximally symmetric space asymptotically are delicate.

\section{Stability issues in the linear equations of motion} \label{sec:stab}

In this work, we focus on spherically symmetric backgrounds, which, without loss of generality, can be written as
\begin{equation}
  \dd s^2 = - A(r) \dd t^2 + \frac{1}{B(r)} \dd r^2 + r^2 \big(\dd \theta^2 + \sin^2\theta \ \dd\varphi^2\big)\,.\label{eq:metric}
\end{equation}
We will introduce the convenient function
\begin{equation}
    Q(r):= \frac{A(r)}{B(r)}\,,
\end{equation}
so that the condition $A=B$ (compatible with ECG) corresponds to having $Q=1$.

An arbitrary perturbation $h_{\mu\nu}$ can be expanded as a multipole series as
\begin{equation}
  h_{\mu\nu} = \sum_{\ell, m} h_{\mu\nu}^\lm\,.\label{eq:hTOhlm}
\end{equation}
If we concentrate in modes with $\ell>1$, we can fix the so-called \emph{Regge-Wheeler gauge}, and, as a result, the corresponding metric perturbation is given in spherical coordinates by the symmetric matrix (we omit the bottom-left part of it):
\renewcommand\arraystretch{1.6}
\begin{equation}
  h^\lm_{\mu\nu}|_ \text{RWg}=\left(\begin{array}{cc|c}
     A(r)H_{0}^\lm(t,r)\Ylm & H_{1}^\lm(t,r)\Ylm & h_{0}^\lm(t,r)\Xlm{}_{a}\\
     \times & \dfrac{1}{B(r)}H_{2}^\lm(t,r)\Ylm & h_{1}^\lm(t,r)\Xlm{}_{a}\\ \hline
     \times & \times & r^2 K^\lm(t,r)\gamma_{ab}\Ylm
   \end{array}\right)\,.
\end{equation}
\renewcommand\arraystretch{1}
See App. \ref{app:gper} for more details on the derivation of this expression and the relevant definitions. We omit the superscripts $\lm$ from now on.

In order to obtain the perturbation equations we proceed as follows. First, we plug the perturbation expansion in the action \eqref{eq:action} and compute the quadratic action $S^{(2)}$, in which perturbations with different $\ell$ decouple, so each of the multipoles can be studied separately. As it is well-known, thanks to the orthogonality properties of the scalar-vector-tensor spherical harmonics and the fact that we are considering a parity-preserving theory, the odd-parity sector, $\{h_0,h_1\}$, and the even-parity one, $\{H_0,H_1,H_2,K\}$, decouple, as well as the different helicities. Under the gauge fixing described above (only valid for $\ell>1$), we focus on the odd-parity sector and derive the equations of motion for $h_0$ and $h_1$, which have the following form:\footnote{Throughout this paper we use the notation $f':= \partial_r f$ and $\dot{f}=\partial_t f$, for any function independent of the angular variables, $f(t,r)$.}
\begin{align}
     {\rm E}_0&:=\frac{8\mpl^{-2}}{\ell(\ell+1)}\frac{\delta S^{(2)}}{\delta h_0} = \mathcal{P}_0 + M_1\ddot{h}_0
    + M_2 h^{\pp}_0
    + M_3\dot{h}^{\p}_1
    + M_4 h^{\p}_0
    + M_5\dot{h}_1
    + M_6 h_0\,, \label{eq:eqh0}\\
     {\rm E}_1&:=\frac{8\mpl^{-2}}{\ell(\ell+1)}\frac{\delta S^{(2)}}{\delta h_1} = \mathcal{P}_1 
    + N_1\dot{h}^{\p}_0
    + N_2\ddot{h}_1
    + N_3 h^{\pp}_1
    + N_4\dot{h}_0
    + N_5 h^{\p}_1
    + N_6 h_1\,,\label{eq:eqh1}
\end{align}
where the objects $\{M_i\}_{i=1}^6$ and $\{N_i\}_{i=0}^6$ are functions of the radial coordinate containing the background functions $A$ and $B$, and their derivatives,\footnote{The highest-order derivatives of $A$ and $B$ can be eliminated by using the background field equations. See App. \ref{app:backg}.} and depend on the multipole order $\ell$ as well through the combination $L := (\ell-1)(\ell+2)$. The objects $\mathcal{P}_0$, $\mathcal{P}_1$, given in \eqref{eq:P0} and \eqref{eq:P1}, correspond to the part of each equation that depends on derivatives of the fields $h_0$ and $h_1$ of order greater than 2.

Interestingly, the system of equations \eqref{eq:eqh0}-\eqref{eq:eqh1} can be simplified by realizing that, in the combination $\dot{{\rm E}}_0+{\rm E}^{\p}_1$, all the fifth-order derivatives cancel and the forth-order part coincides with the one of ${\rm E}_1$ but multiplied by the factor $\frac{2}{r}+\frac{Q'}{2Q}$. Therefore under the condition $\frac{2}{r}+\frac{Q'}{2Q}\neq 0$ (which is true in particular when $A=B$), the system $\{{\rm E}_0=0,{\rm E}_1=0\}$ is equivalent to $\{{\rm E}_0=0,\tilde{{\rm E}}_1=0\}$, with
\begin{equation}
    \tilde{{\rm E}}_1 := \left(\frac{2}{r}+\frac{Q'}{2Q}\right){\rm E}_1+\dot{{\rm E}}_0+{\rm E}^{\p}_1\,.
\end{equation}
The structure of this new equation is:
\begin{equation}
    \tilde{{\rm E}}_1 = \tilde{\mathcal{P}}_1 +O_1\dot{h}^{\p}_0+ O_2 h^{\pp}_1+ O_3\ddot{h}_1 + O_4\dot{h}_0+ O_5h^{\p}_1 +O_6h_1,\label{eq:eqh1t}
\end{equation}
where the radial functions $\{O_i\}^6_{i=1}$ have similar dependencies as $M_i$ above and $\tilde{\mathcal{P}}_1$, which is given in \eqref{eq:P1t}, is the piece depending on third-order derivatives of $h_0$ and $h_1$. The resulting equations could be further simplified and put in normal form, but that is not necessary for our purposes here and the form \eqref{eq:P1t} will be sufficient. The first property to notice is that the system contains up to third-order time derivatives of $h_0$ and $h_1$. This means that, in general, we will need up to six initial conditions, thus signaling the presence 3 Lagrangian dof's, in agreement with the results of \cite{DeFelice2023} that find the same number of dof's with an alternative analysis. The remaining five dof's of the generic cubic theory will live in the even sector. Notice that the presence of three propagating dof's in the odd sector already allows to conclude that this sector will unavoidably have a ghost without having to explicitly compute the corresponding kinetic matrix, a conclusion also supported by the explicit computation of \cite{DeFelice2023}.

Let us now analyse the highest-order derivative (i.e. the principal) part of these equations. The full expressions are collected in App. \ref{app:hosectors}. Here we just focus on the parts with higher-than-2nd-order time derivatives, which are totally characterised by two functions $C_0$ (in ${\rm E}_0$ and ${\rm E}_1$) and $C_1$ (in $\tilde{{\rm E}}_1$):
\begin{align}
    \mathcal{P}_0 &= C_0 \dddot{h}^{\p}_1+\left[C^{\p}_0 +\left(\frac{2}{r}+\frac{Q'}{2Q}\right) C_0\right]\dddot{h}_1 
    +  (\text{lower order in time der.}) \,,\\
    \mathcal{P}_1 &= -C_0 \left(\ddddot{h}_1-\dddot{h}^{\prime}_0 \right) -\frac{2}{r}C_0 \dddot{h}_0 +  (\text{lower order in time der.}) \,,\\
    \tilde{\mathcal{P}}_1 &= C_1 \dddot{h}_0 + (\text{lower order in time der.}) \,.
\end{align}
If we include the terms that we are omitting here (see \eqref{eq:P0}-\eqref{eq:P1t}), it turns out that all the coefficients of $\mathcal{P}_0$ and $\mathcal{P}_1$ can be written in terms of $C_0$ and $C^{\p}_0$ plus corrections that go with either $Q'$ or $Q''$. Similarly, all the coefficients of $\tilde{\mathcal{P}}_1$ go either with $C_1$ or with radial derivatives of $Q$. The explicit expressions of these relevant functions are:
\begin{align}
    C_0 &:= \frac{\bECG}{\cutoff^4}\frac{3}{rQ^2} \left[\left(1-2\aECG+(\gamma-1)\frac{r A'}{2A}\right)\frac{Q'}{Q}-\frac{2}{r}\left(1-\frac{1}{B}+(\gamma-2)\frac{rA'}{2A}+(\gamma-1)\frac{r^2A''}{2A}\right)\right]\,,\label{eq:C0}\\
    C_1 &:= \frac{\bECG}{\cutoff^4}\frac{6 L}{r^3 AQ} \left[\left(1+(\aECG-1)\frac{r A'}{2A}\right)\frac{Q'}{Q}-\frac{2}{r}\left(-\aECG\left(1-\frac{1}{B}\right)+\gamma\frac{rA'}{2A}+(\aECG-1)\frac{r^2A''}{2A}\right)\right]\,.\label{eq:C1}
\end{align}
In particular, for $B=A$ ($Q=1$), we find:
\begin{align}
    C_0\big|_{B=A} &= -\frac{\bECG}{\cutoff^4}\frac{6}{r^2} \left[1-\frac{1}{A}+(\gamma-2)\frac{rA'}{2A}+(\gamma-1)\frac{r^2A''}{2A}\right]\,,\\
    C_1\big|_{B=A} &= -\frac{\bECG}{\cutoff^4}\frac{12 L}{r^4 A} \left[-\aECG\left(1-\frac{1}{A}\right)+\gamma\frac{rA'}{2A}+(\aECG-1)\frac{r^2A''}{2A}\right]\,.
\end{align}

Let us first corroborate that we recover the known results that the ECG has a reduced linear spectrum with respect to the general cubic case around maximally symmetric backgrounds. Thus, we will assume a background function of the form $A(r)=B(r) = 1 - a r^2 $ that describes Minkowski, de Sitter and anti-de Sitter for vanishing, positive and negative $a$ respectively. The expressions of $C_0$, $C^{\p}_0$ and $C_1$ become substantially simpler and can be written as
\begin{align}
    C_0(r)|_{\text{MaxSym}}&= 12(\gamma-1)\frac{\bECG}{\cutoff^4} \frac{a}{1-ar^2}\,, \\  
    C^{\p}_0(r)|_{\text{MaxSym}}&= 24(\gamma-1)\frac{\bECG}{\cutoff^4} \frac{r a^2}{(1-ar^2)^2}\,,  \\ 
    C_1(r)|_{\text{MaxSym}}&= 12 L (\gamma -1)\frac{\bECG}{\cutoff^4}  \frac{a}{r^2(1-a r^2)^2}\,.
\end{align}
There are two important observations worth making at this point. Firstly, Minkowski spacetime ($a=0$) has vanishing $C_0$, $C^\p_0$ and $C_1$ which means that, regardless the value of $\gamma$, Minkowski will always exhibit fewer dof's at linear order. This is however expected since Minkowski space has vanishing curvature so the Riemann tensor starts at first order in perturbations and, hence, the cubic term does not contribute to the quadratic action for the perturbations (see eq. \eqref{eq:pertMaxSym} for $\mathcal{R}=0$). The second important observation is that all the relevant coefficients that govern the principal part of the perturbation equations are proportional to $\gamma-1$ so we recover the reduced linear spectrum of the ECG around maximally symmetric spacetimes. This result is independent of $\aECG$. 

Let us now proceed to discuss how this singular behaviour of the maximally symmetric solutions transcends to spherically symmetric backgrounds with an asymptotic region is maximally symmetric. Consider an arbitrary initial configuration that deviates from a certain asymptotically flat solution by a generic (but small) perturbation. Let us focus on how this initial condition evolves very far from the origin, where the spacetime is flat in good approximation.  In a generic initial configuration of this kind, there will be active modes that are strongly coupled as we take the limit of zero curvature in the background. Although our spherically symmetric background has non-vanishing curvature, the previously mentioned modes are sensitive to the flatness of the asymptotic region. During the time evolution, these modes will start growing asymptotically, as they would do in Minkowski space, and the generated solution will either deviate from the unperturbed one or reach a singular behaviour (e.g. infinite derivative at a finite time). This is a similar situation as the one described in the previous publication \cite{BeltranJC2021}, but happening only in the region very far from the origin. This is actually generic for any cubic theory (not only \eqref{eq:action}). The same conclusions apply to any asymptotically (A)dS solution of \eqref{eq:action} with $\gamma = 1$ (see also \eqref{eq:pertMaxSym}). However, for $\gamma \neq 1$ the strong coupling issue disappears but then all the dof's (including the ghostly ones) are fully active, thus casting doubts on the physical viability of these solutions.

\section{Discussion} \label{sec:conclusions}

In this note, we have considered the viability of black hole solutions in gravity theories featuring higher-order curvature terms, focusing for the sake of generality in the cubic terms. Our motivation for this analysis has been some works in the literature that study modified black hole scenarios based on the ECG as well as other specific cubic theories with reduced linear spectra around maximally symmetric backgrounds. These theories can be interpreted in two different ways, as full theories on their own or as the next-to-leading order correction to GR without matter as an EFT. In the latter, these terms can only contribute perturbative corrections to the GR predictions both in the background solutions and in the perturbations, so no particular jeopardy is found here, as explained in detail in \cite{Bueno2023}. On the other hand, if we want to have non-perturbative corrections to the GR behaviour, as required for instance to modify the horizon or to resolve the singularities, then these effects come hand in hand with pathologies. In an attempt to alleviate these pathologies, the cubic interactions can be tuned to remove the additional dof's around maximally symmetric backgrounds that would conform the asymptotic behavior of the found black hole solutions. Our main message here is that this mechanism does not resolve the problem, but it only hides it by arguably making it even worse with additional strong coupling problems. Thus, either the effects are perturbative and potential instabilities as those discussed in \cite{DeFelice2023} are harmless within the regime of validity of the EFT, or the effects are non-perturbative but then the ghosts will make it physically non-viable. Let us emphasize however that these theories, despite their pathological character when regarded as full theories and not in the framework of EFTs, exhibit interesting properties like the possibility of analytically computing thermodynamical quantities of black holes and the relation to holography \cite{Bueno2017BHstabil,Myers2010b}.

It might be worth mentioning that, among the cubic theories there is one particular case where one can have non-perturbative effects while taming the ghost if, among the cubic terms, there is a hierarchy between the $R^3$ and the other operators. In this case, the $R^3$ term describes one additional scalar mode that can come in at a lower scale than the ghost.

To finish, we shall stress once again that our findings, though obtained for the cubic theories, straightforwardly apply to all other higher-order curvature extensions featuring a reduced linear spectrum around specific backgrounds, barring the known cases such as $f(R)$ where the reduced spectrum is a property of the full theory.

\begin{acknowledgments}
The authors would like to thank Gerardo Garc\'ia Moreno, Jose Alberto R. Cembranos, Pablo A. Cano and Shinji Tsujikawa for useful comments and feedback. JBJ is supported by Project PID2021-122938NB-I00 funded by the Spanish “Ministerio de Ciencia e Innovaci\'on” and FEDER “A way of making Europe”. AJC is is supported by the European Regional Development Fund through the Center of Excellence TK133 “The Dark Side of the Universe” and the Mobilitas Pluss post-doctoral grant MOBJD1035. 
\end{acknowledgments}

\appendix

\section{Spherical harmonics} \label{app:harmonics}

We introduce the following Hilbert space product for $\mathbb{C}$-valued square-integrable covariant tensors on the sphere:
\begin{equation}
  \left\langle T_{a...b}, S_{c...d} \right\rangle := \sphint{\gamma^{ac}...\gamma^{bd} (T_{a...b})^{*} S_{c...d}} 
  = \sphint{(T_{a...b})^{*} S^{a...b}}\,. \label{eq:scalarprod}
\end{equation}
If we restrict to $\mathbb{R}$-valued tensors, the complex conjugate drops and we get just the integral of the contraction of the two tensors (with the $\mathbb{S}^2$ metric).

If we call $\{\mathcal{Y}_\lm\}$ a basis of complex spherical harmonics normalised as $\left\langle \mathcal{Y}_\lm, \mathcal{Y}_\lmp \right\rangle= \delta_{\ell \ell'}\delta_{mm'}$, the real basis $\{\Ylm\}$ we are using in this manuscript  is constructed as follows:
\begin{equation}
  \Ylm(\theta,\varphi):=\left\{ \begin{array}{lrl}
\mathcal{Y}_{\ell0} & \text{if}\  & m=0\\
\sqrt{2}\mathrm{Re}\mathcal{Y}_{\ell|m|} & \text{if}\  & m>0\\
\sqrt{2}\mathrm{Im}\mathcal{Y}_{\ell|m|} & \text{if}\ & m<0
\end{array}\right.=\left\{ \begin{array}{lrl}
\mathcal{Y}_{\ell0} & \text{if}\ & m=0\\
\frac{1}{\sqrt{2}}\Big[\mathcal{Y}_\lm+(\mathcal{Y}_\lm)^{*}\Big] & \text{if}\  & m>0\\
\frac{1}{\iN\sqrt{2}}\Big[\mathcal{Y}_{\ell|m|}-(\mathcal{Y}_{\ell|m|})^{*}\Big] & \text{if}\  &m <0
\end{array}\right.
\end{equation}
These harmonics satisfy the analogous eigenvalue and normalization conditions of their complex counterparts:
\begin{align}
\left[ \gamma^{ab} D_a D_b   + \ell(\ell+1)\right] Y_\lm&=0\,,\label{eq:HelmY}\\
  \left(\partial_\varphi^2+m^2\right)Y_\lm &=0, \\
  \left\langle \Ylm, Y_\lmp \right\rangle&= \sphint{\Ylm Y_\lmp} 
   =\delta_{\ell \ell'}\delta_{mm'}\,.
\end{align}

\section{Metric perturbations and gauge fixing} \label{app:gper}

The general spherically symmetric metric \eqref{eq:metric} can be also be written
\begin{equation}
  \dd s^2 = \bar{g}_{\mu\nu} \dd x^\mu \dd x^\nu = f_{AB}(z)\dd z^A \dd z^B + r^2(z) \gamma_{ab}(\vartheta) \dd \vartheta^a \dd \vartheta^b\,,
\end{equation}
where $\gamma_{ab}$ is the metric of the 2-sphere and $f_{AB}$ covers the remaining two directions. In standard spherical coordinates $z^A=\{t,r\}$, $\vartheta^a=\{\theta, \varphi\}$, we recover \eqref{eq:metric}.

The arbitrary perturbation \eqref{eq:hTOhlm} can be split into the so-called scalar, vector and tensor sectors:
\begin{align}
h_{AB}^\lm & =S_{AB}^\lm(t,r)\Ylm \,,\\
h_{Aa}^\lm & =V_{A}^\lm(t,r)\Ylm{}_{a}+h_{A}^\lm(t,r)\Xlm{}_{a}\,,\\
h_{ab}^\lm & = r^2 K^\lm(t,r)\gamma_{ab}\Ylm+G^\lm(t,r)\Ylm{}_{ab}+E^\lm(t,r)\Xlm{}_{ab}\,.
\end{align}
Here we are using a basis of real scalar spherical harmonics $\{\Ylm\}$ (see App. \ref{app:harmonics}) and the vector and tensor spherical harmonics are given by:
\begin{align}
    \Ylm{}_a &:= D_a \Ylm\,,\\
    \Xlm{}_a &:= \gamma^{bc} \varepsilon_{ab}  D_c \Ylm\,,\\
    \Ylm{}_{ab} &:= \left[ D_a D_b + \frac{1}{2}\ell(\ell+1)\gamma_{ab}\right] \Ylm\,,\\
    \Xlm{}_{ab} &:= -\gamma^{cd}\varepsilon_{ca} D_{b}D_d \Ylm\,,
\end{align}
where $D_a$ and $\varepsilon_{ab}$ are respectively the covariant derivative and the Levi-Civita tensor associated to the metric of the 2-sphere, $\gamma_{ab}$.

In the scalar sector we use the notation:
\begin{equation}
S_{tt}^\lm =: A(r)H_{0}^\lm(t,r)\,,\qquad S_{tr}^\lm=: H_{1}^\lm(t,r)\,,\qquad S_{rr}^\lm=:\frac{1}{B(r)}H_{2}^\lm(t,r)\,.
\end{equation}
Notice that all the time and radial functional dependence has been encoded in 10 functions, which are scalars under rotations,
\begin{equation}
    \{H_{0}^\lm ,\ H_{1}^\lm ,\ H_{2}^\lm ,\  V_{0}^\lm ,\ V_{1}^\lm ,\ K^\lm ,\ G^\lm\},\qquad \{ h_{0}^\lm ,\ h_{1}^\lm ,\ E^\lm \}.
\end{equation}
The functions in the first set correspond to the even-parity sector of perturbations and the others to the odd-parity one.

Some of these functions can be removed thanks to the gauge symmetry. Under an infinitesimal coordinate transformation $x^\mu\to x^\mu +\xi^\mu$, the metric perturbation transforms as
\begin{equation}
    \delta h_{\mu\nu} = - 2 \nabla_{(\mu}\xi_{\nu)}.
\end{equation}
The co-vector $\xi_{\mu}$, as the metric, can be expanded in terms of scalar and vector spherical harmonics:
\begin{equation}
  \boldsymbol{\xi} = \sum_{\ell,m} \left[\zeta^\lm_A(t,r)\Ylm\right] \dd z^A   + \sum_{\ell,m} \left[\Pi^\lm(t,r)\Ylm{}_{a}+\alpha^\lm(t,r)\Xlm{}_{a}\right] \dd y^a\,.\label{eq:xitolm}
\end{equation}
By using this decomposition, the transformation rules for the different perturbation functions can be written as
\begin{align}
\delta H_0 & =  -\frac{1}{A} \left(2 \dot{\zeta}_0- A' B \zeta_1 \right) \,,
&\delta K   & = \frac{1}{r^2} \left[  \ell(\ell+1)\Pi - 2Br \zeta_1\right] \,, \nonumber\\
\delta H_1 & =  - \zeta^\prime_0 - \dot{\zeta}_1 + \frac{A'}{A} \zeta_0 \,, 
&\delta G   & = -2 \Pi \,,\nonumber\\
\delta H_2 & = -B \left(2  \zeta^\prime_1+ \frac{B'}{B} \zeta_1 \right) \,,
&\delta E   & = -2\alpha \,,\nonumber\\
\delta V_0 & = -\zeta_0 - \dot{\Pi}  \,,
&\delta h_0 & = -\dot{\alpha}\,,\nonumber\\
\delta V_1 & = -\zeta_1 + \frac{2}{r} \Pi  -\Pi^\prime \,,
&\delta h_1 & = - \alpha^\prime + \frac{2}{r}\alpha\,.
\end{align}
where we have dropped the superscript $\lm$ to alleviate notation. 

The Regge-Wheeler gauge mentioned in the main text corresponds to the choice
\begin{equation}
    \text{RWg}:=\{V^\lm_0 = V^\lm_1 = G^\lm = E^\lm = 0\}\,,
\end{equation}
which can always be done for $\ell>1$.

\section{Noether identity for spherically symmetric spaces} \label{app:Noetherid}

For the case of a theory depending exclusively on the metric tensor, the Noether (also known as generalised Bianchi) identity associated to diffeomorphims tell us that
\begin{equation}
    \nabla_\mu \EoM^{\mu\nu} = 0.
\end{equation}
Now we expand the covariant derivative by using the Christoffel symbols of the static spherically symmetric metric \eqref{eq:metric} and use that there are only diagonal contributions to $\EoM^{\mu\nu}$, and the result is
\begin{align}
    \partial_t \EoM^{tt}&=0\,,\\
    \left[\partial_r +\left(\frac{2}{r}+\frac{A'}{2A}-\frac{B'}{B}\right) \right]\EoM^{rr}+\frac{A'B}{2}\EoM^{tt}-rB\left(\EoM^{\theta\theta} +\sin^2\theta\EoM^{\varphi\varphi}\right) &=0\,,\\
     \partial_\theta\EoM^{\theta\theta}+\cot\theta\,\EoM^{\theta\theta} -\sin\theta\,\cos\theta\,\EoM^{\varphi\varphi}&=0\,,\\
    \partial_\varphi \EoM^{\varphi\varphi}&=0\,.
\end{align}
Since $\partial_\varphi \EoM^{\varphi\varphi}=\partial_t \EoM^{tt}=\partial_\theta\EoM^{\theta\theta}=0$, we arrive at the system of equations:
\begin{align}
    \left[\partial_r +\left(\frac{2}{r}+\frac{A'}{2A}-\frac{B'}{B}\right) \right]\EoM^{rr}+\frac{A'B}{2}\EoM^{tt}-2rB\EoM^{\theta\theta} &=0\,,\\
     \EoM^{\theta\theta} -\sin^2\theta\,\EoM^{\varphi\varphi}&=0\,.
\end{align}
Here we clearly see that it is enough to solve the equations $\EoM^{tt}=0$ and $\EoM^{rr}=0$, because the angular ones will be automatically satisfied by means of the Noether identity.

\section{Background equations}
\label{app:backg}

The equations of motion of the theory are given by
\begin{equation}
    0=\EoM^{\mu\nu}:=\frac{2}{\mpl^2}\frac{1}{\sqrt{|g|}}\frac{\delta S}{\delta g_{\mu\nu}}\,.
\end{equation}
As it is well known, by virtue of the Noether identity associated to diffeomorphisms (see App. \ref{app:Noetherid}), for metrics of the type \eqref{eq:metric}, one just need to solve the $tt$ and $rr$ components of the equation of the metric, which we multiply by appropriate factors for convenience,
\begin{equation}
    \EoM_1:=4A(r) \EoM^{tt}\,,\qquad \EoM_2:=\frac{4}{B(r)}\EoM^{rr}\,.
\end{equation}

Let us  focus on maximally symmetric backgrounds: Minkowski spacetime is a trivial solution for all theories \eqref{eq:action} with $\Lambda_0=0$, whereas a de Sitter spacetime $B(r)=A(r)=1-\mathcal{R} r^2$ requires the cubic equation \eqref{eq:bgeqMaxSym} to be satisfied. 

\subsection*{Derivative content}

If we check the derivative content of these equations, we find that $\EoM_1$ depends up to $A''''$ and $B'''$, whereas $\EoM_2$ is of a lower order, depending up to $A'''$ and $B''$. Therefore, one can use the derivative of the second one to eliminate the fourth order derivative of $A$ in the first equation. At the end one arrives at a system of two third-order ordinary differential equations for $A$ and $B$. 

The second equation has the form
\begin{equation}
    \EoM_2 = F(r,A,A',A'',B,B') A''' + G(r,A,A',A'',B,B',B'')\,.
\end{equation}
This allows, as mentioned above, to solve for $A'''$. However, this is only possible if $F$ is non-vanishing. This function is given by the following expression
\begin{align}
    F(r,A,A',A'',B,B') &:= \frac{3}{4
   r^3}\left(\frac{B}{A}\right)^3\frac{\bECG}{\cutoff^4}\Bigg[(\gamma-1)  r^3A' \left(2  A'' - A' \frac{B}{A} \left(\frac{A}{B}\right)'\right)\nonumber\\
   &\qquad\qquad\qquad +4r A A'-8A^2\left(1-\frac{1}{B}\right) + 8 (\aECG-1)r A B \left(\frac{A}{B}\right)'\Bigg]\,.
\end{align}

We are interested in exploring the cases around $B=A$ and $\gamma=1$. If we impose the former, it can be shown that the parameter $\aECG$ disappears, implying that the higher-order contribution coming from ECG and its cosmological improvement are indistinguishable for this kind of backgrounds. In other words, the Lagrangian density $\mathcal{C}$ gives no contribution to the background equations. The function $F$ in this case is given by: 
\begin{equation}
    F(r,A,A',A'',A,A') = \frac{3}{2
   r^3}\frac{\bECG}{\cutoff^4}\Big[(\gamma-1)  r^3A'A'' +2A(rA'+2)-4A^2\Big],
\end{equation}
so we clearly see that the more extreme case $\{B=A, \gamma=1\}$ gives an $F$ that is not identically vanishing. Therefore, generically, one can use the equations of motion of the background to eliminate $A'''$ and $B'''$ (and higher derivatives) from the perturbation equations.

\section{Highest-order part of the odd-parity equations}
\label{app:hosectors}

The highest-order derivative pieces of the odd-parity perturbation equations \eqref{eq:eqh0}, \eqref{eq:eqh1} and \eqref{eq:eqh1t} read:
\begin{align}
    \mathcal{P}_0 &= C_0( \dddot{h}^{\p}_1- \ddot{h}^{\pp}_0)+ AB\left(C_0+\mu \frac{Q'}{Q}\right)(h_0^{\pppp}-\dot{h}^{\ppp}_1) \nonumber\\
    &\quad - \left(C^{\p}_0 +\frac{Q'}{2Q} C_0\right) \ddot{h}^{\p}_0 +\left[C^{\p}_0 +\left(\frac{2}{r}+\frac{Q'}{2Q}\right) C_0\right]\dddot{h}_1 
    +2 \zeta(1)h^{\ppp}_0      
    - 2\zeta(0) \dot{h}^{\pp}_1 \,,\label{eq:P0}\\
    \mathcal{P}_1 &= -C_0 (\ddddot{h}_1-\dddot{h}^{\prime}_0 ) + AB\left(C_0+\mu \frac{Q'}{Q}\right)(\ddot{h}^{\pp}_1-\dot{h}^{\ppp}_0)
    \nonumber\\
    &\quad -\frac{2}{r}C_0 \dddot{h}_0
    +  \zeta(1)\ddot{h}^{\p}_1- \zeta(3)\dot{h}^{\pp}_0  \,,\label{eq:P1}\\
    \tilde{\mathcal{P}}_1 &= C_1 \dddot{h}_0 - \varrho(\gamma+4\aECG-4)\dot{h}^{\pp}_0-\varrho(\gamma-4\aECG)\ddot{h}^{\p}_1+ AB \varrho(2\gamma-4) h^{\ppp}_1 \,, \label{eq:P1t}
\end{align}
where, for each equation, we are separating in different lines the fourth- and the third-order sectors. The functions $C_0$ and $C_1$ are given in \eqref{eq:C0}-\eqref{eq:C1}. Here we are using the following abbreviations:
\begin{align}
    \mu&:=3(\gamma+4\aECG-4) \frac{\bECG}{\cutoff^4}\frac{1}{r Q^2}\,,\\
    \zeta(j)&:=AB \left\{ C^{\p}_0 +\left(\frac{1-j}{r}+\frac{2 A'}{A}-\frac{Q'}{2Q}\right) C_0 + \mu \left[\left(\frac{2A'}{A}-\frac{j}{r}\right)\frac{Q'}{Q} - \frac{7}{2}\left(\frac{Q'}{Q}\right)^2 + \frac{Q''}{Q}\right]\right\} \,,\\
    \varrho(j)&:= AB \left(C_1+3 L j \frac{\bECG}{\cutoff^4} \frac{Q'}{r^3 A Q^2}\right)\,.
\end{align}

\bibliographystyle{JHEP}
\bibliography{bibliography.bib}

\end{document}